# Native Chemical Automata and the Thermodynamic Interpretation of Their Experimental Accept/Reject Responses[*]

Marta Dueñas-Díez[†] and Juan Pérez-Mercader[‡]

‡ To whom correspondence should be addressed.

## *Introduction*

Computation—defined as the pathway for information to be input, to be processed mechanically, and to be output in a useful way (Evans 2011)—takes place not only in the myriad of electronic devices we use daily but also in living systems. Life carries out computations mostly by using chemical support: inputs are chemical substances, the mechanical processing occurs via chemical reaction mechanisms, and the result is chemical as well. Machines carrying out computations are typically referred to as automata (Hopcroft, Motwani, and Ullman 2006); hence, to a large extent, living systems can be viewed as chemical automata (Bray 2009). Classic automata are arranged hierarchically from simplest to most powerful (Hopcroft, Motwani, and Ullman 2006): finite automata, then pushdown automata, and, at the top of the hierarchy, Turing machines (Turing 1936).

---

[†]Repsol Technology Center, Carretera de Extremadura S/N, 28935 Móstoles, Madrid, Spain; Department of Earth and Planetary Sciences and Harvard Origins of Life Initiative, Harvard University, 20 Oxford Street, Cambridge, Massachusetts 02138; martaduenasdiez@fas.harvard.edu

[‡]Department of Earth and Planetary Sciences and Harvard Origins of Life Initiative, Harvard University, 20 Oxford Street, Cambridge, Massachusetts 02138; Santa Fe Institute, 1399 Hyde Park Road, Santa Fe, New Mexico 87501; jperezmercader@fas.harvard.edu

Although the subject of this contribution already has an interesting history, we give here a brief, personal, and short summary of some interesting developments in the field of chemical computation. Interest in chemical computing dates back to the early 1970s, when Conrad (1972) studied information processing in molecular systems and how it differs from electronic digital computing. A theoretical chemical diode was first suggested by Okamoto, Sakai, and Hayashi (1987), an idea that Hjelmfelt, Weinberger, and Ross (1991) further developed to suggest that neural networks and chemical automata could be constructed connecting such chemical diodes. In the 1990s, Magnasco (1997) studied the Turing completeness of chemical kinetics. The first experimental realization of chemical AND and OR logic gates using reaction diffusion was achieved in 1995 by Tóth and Showalter (1995), followed by XOR gates (Adamatzky and Lacy Costello 2002) and counters (Górecki, Yoshikawa, and Igareshi 2003), and still is an active area of research due to the difficulties associated to linking many gates to carry out more advanced computations. Computations carried out in a more native way, without requiring diffusion, have been suggested using complex biomolecules such as DNA (Adleman 1994; Benenson 2009) or chromatin (Prohaska, Stadler, and Krakauer 2010; Bryant 2012). In summary, most artificial approaches to chemical computing, inspired by living systems, focus on reaction–diffusion systems mostly representing logic gates or use complex biomolecules to solve very specific problems.

Our approach (Pérez-Mercader, Dueñas-Díez, and Case 2017) differs from the aforementioned work in that we use the power of chemistry, and the molecular recognition associated with the occurrence of chemical re-actions, in a one-pot reactor, that is, a single well-mixed container where multiple rounds of reactions can take place, without using external geometrical aids or complex biomolecules and relying fully on the power of molecular recognition and the robustness associated with Avogadro's number to carry out computations. We have recently demonstrated experimentally that this approach, without using biochemistry, can recognize a language that only automata at the Turing machine level of the hierarchy can recognize (Dueñas-Díez and Pérez-Mercader, Submitted 2018).

In this contribution, we apply the well-known natural connection between chemistry and thermodynamics (Donder 1927; Kondepudi and Prigogine 2014) to study and interpret the chemical reject/accept signatures of chemical automata in thermodynamic terms. We do this for three examples, one at each main level of the three-level hierarchy in computing automata theory (Hopcroft, Motwani, and Ullman 2006). Of course, this connection is only a first step toward quantifying the thermodynamic cost of chemical computation and, more importantly, toward its optimization (Bennett 1982; Landauer 1961). Indeed, the same thermodynamic metrics we apply as the reject/accept signatures after a word is processed can be applied during the course of the computation, not just at its end. If we apply our metrics continuously during the computation of a complete sequence, we can assess the thermodynamic cost of computation as each symbol is processed and therefore determine how the thermodynamic cost evolves as the input sequence length grows or even compare the cost of different types of rejects. We suggest using the three languages chosen below, or other similar well-known languages and their associated automata, as minimal complete examples to run quantitative studies of the thermodynamic cost of computation.

## *One-Pot Native Chemical Computation*

Our work focuses on demonstrating experimentally how computations of different complexity (in the sense of classical automata theory) can be carried out by chemical means exclusively, in a homogeneous reactor, and without requiring complex biomolecules. In our approach, the input to be computed is represented by a sequence of symbols from a chemical alphabet in which each letter corresponds to a certain constant amount, or aliquot, of a carefully chosen reacting chemical species. The input is sequentially added letter by letter, that is, aliquot by aliquot, to a one-pot reactor at constant time intervals (Pérez-Mercader, Dueñas-Díez, and Case 2017). The processing of each letter consists in selectively activating specific pathways in the chemical mechanism and, correspondingly, altering the resulting chemical state/landscape in a systematic way. Finally, the output of the computation is in the form of a distinct chemical response; that is, for a given automata/language combination, the chemical behavior associated with a rejected sequence is different from the chemical behavior associated to that of an accepted sequence (Dueñas-Díez and Pérez-Mercader, Submitted 2018). Naturally, we expect that such distinct chemical responses correspond to some distinct thermodynamic signatures as well.

To show that chemistry can carry out computations of different complexity levels, we carry out the following steps. First, we choose a specific language of interest that a tailored chemical automaton should recognize, that is, a specific problem to be solved. From classic automata theory, we then identify the class of automata needed to recognize it and the computational requirements as defined by the corresponding automata tuple. We translate this into specific requirements for the chemical reactions and their reactants, products, and intermediates, leading us to select the alphabet description appropriately. Then, the specific quantitative recipes for initial conditions and alphabet aliquots are selected so that the chemical reaction monitoring system allows detection of the machine's response with sufficient/reasonable precision.

## *Chemical Finite Automaton Recognizing the Regular Language $L_1$ of All Words Containing at Least One "A" and One "B"*

Regular languages are the simplest languages in automata hierarchy (Hopcroft, Motwani, and Ullman 2006). They do not require counting: their words all contain or all exclude certain patterns of the alphabet symbols or affixes (Hopcroft, Motwani, and Ullman 2006; Cohen 1991). Regular languages are recognized by a finite automaton (FA), an abstract device that at each given time is in one of a finite number of states, and the device transitions states depending on the input using a finite set of rules. At the end of a computation, that is, after a word is processed by the FA, the device terminates in either an accept state or a reject state, depending on whether the input word belongs to the regular language recognized by the FA.

Following the intuitive notion that simple chemistries can recognize regular languages, we reverse-formulate the question as follows: what can a single bimolecular reaction of the type $A + B \rightarrow C + D$ compute? If we represent the letters, **a** and **b**, in a language $L_1$ by the aliquots of $A$ and $B$ corresponding to this reaction, we see that such a bimolecular reaction recognizes the regular language $L_1$ of all words that contain at least one **a** and one **b**. For an illustrative and visual implementation, we can choose a precipitation reaction in an aqueous medium such as

$$\text{KIO}_3 + \text{AgNO}_3 \rightarrow \text{AgIO}_3(s) + \text{KNO}_3. \tag{1}$$

If, during computation, a white precipitate of silver iodate is observed, the input string has been recognized and accepted; if the solution is clear from precipitate, the string has been rejected because there was no reaction and the input string was therefore not recognized. The only requirement in this example to choose the recipes of alphabet symbols **a** (potassium iodate) and **b** (silver nitrate) quantitatively is that the product of their concentrations, once one aliquot of each is added to the reactor, exceeds the solubility product constant of silver iodate at the operating reactor temperature, thus guaranteeing the appearance of a precipitate. Fig. 1 shows the chemical representation of symbols **a** and **b**, the bimolecular precipitation reaction, the corresponding theoretical FA transition graph to recognize $L_1$, and the results of testing two sequences experimentally. Sequence *aab* gives a white precipitate, corresponding to the final state $q_f$ in the abstract FA. Sequence *aaa* shows no precipitate, corresponding to state $q_1$, and thus the input is rejected by the chemical automaton.

Because this reaction is exothermic, the accept and reject states can also be detected by monitoring temperature (if the temperature remains constant, the sequence is rejected, but if the temperature increases, then the sequence is accepted). Hence we see that the heat of reaction is the thermodynamic equivalent to the chemical precipitate response.

### *Chemical 1-Stack PDA Recognizing the Context-Free Language $L_2$ of Balanced Parentheses*

Next, we go one important step up in the hierarchy and consider the case of context-free languages. We show how one-pot native chemistry recognizes a language in which both counting and sequence order are relevant. Context-free languages (CFL) are those whose words involve matching of substrings, affixes, or symbols and therefore require counting to one arbitrarily high integer (Hopcroft, Motwani, and Ullman 2006). We choose the Dyck language (Weisstein 2009), the language of balanced parentheses, as $L_2$: a sequence of parentheses is balanced if, during its processing, the number of closed parentheses never exceeds the number of open parentheses, and at the end of the computation, the number of open parentheses matches exactly the number of closed parentheses (Hopcroft, Motwani, and Ullman 2006; Cohen 1991).

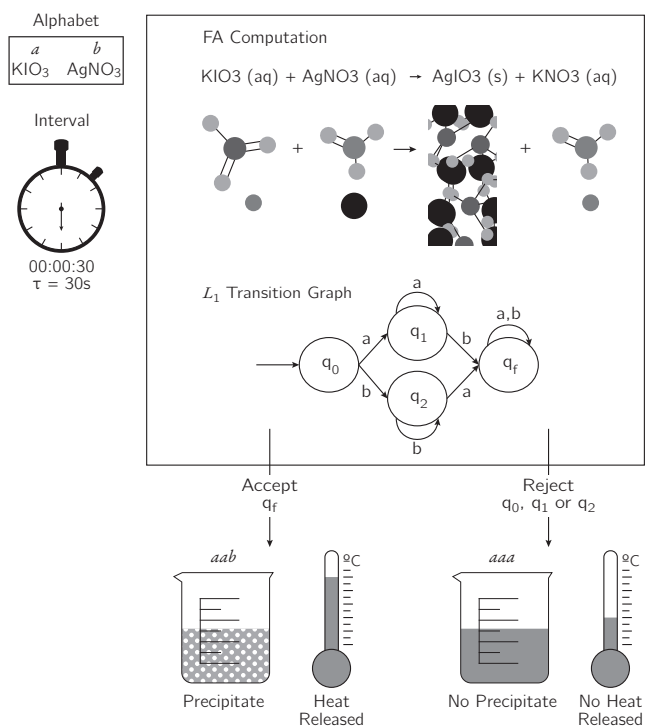

**Figure 1.** *Operation of a chemical finite-state automaton: language $L_1$, described by the regular expression $(a+b)^* \, a(a+b)^*b(a+b)^* + bb^* \, aa^*$, is recognized by a FA and is realized chemically by a precipitation reaction. In this example, once the full sequence has been processed, if the solution contains a visible precipitate of $AgIO_3$, or, equivalently, heat has been released during computation, the input string has been accepted as a word in $L_1$. In contrast, if the solution does not contain any visible precipitate, or, equivalently, no heat was released during computation, then the input string was rejected. Input sequences aab and aaa were tested experimentally and in the former a precipitate was observed, while no precipitate was observed in the latter.*

In theoretical computer science, a CFL is recognized by a one-stack pushdown automaton (PDA). This automaton differs from a FA by being endowed with an additional element, the stack, in which to store a string of arbitrary length and that, furthermore, can be read and modified only at its top, in a last-in-first-out fashion (Hopcroft, Motwani, and Ullman 2006; Cohen 1991), just as in a cafeteria "stack of trays." The transitions in a PDA depend not only on the current input symbol and state but also on the current symbol at the top of the stack. A transition may result not only in changing the state of the automaton but also in pushing (adding) an element to the top of the stack or popping (removing) an element from the stack. The "accept" criterion is often associated with the set of transitions leading to an empty stack at the end of the computation. For our chosen language $L_2$, the stack keeps track of the excess of open parentheses with respect to the closed parentheses and, indeed, it has to be empty at the end of the computation.

The requirement of a "stack" translates in native chemical computing into the condition of having a pathway in the reaction mechanism in which there is an intermediate species that is produced (pushed) in one subreaction and consumed (popped) in another subreaction. This in turn leads us to select as an example for the actual implementation of a chemical $L_2$-PDA the language of Dyck words by means of a pH reaction and with the following alphabet assignment: "(" is an aliquot of the base (NaOH), ")" is an aliquot of the weak diprotic acid ($CH_2(COOH)_2$), and "#"—the symbol that delimits the beginning- and end - of-sequence— is an aliquot of a $pH$ indicator. The quantification of the recipes of the symbol aliquots is carried out so that one aliquot of "(" and one aliquot of ")" neutralize each other to the midpoint in the $pH$ curve (Petrucci et al. 2011) and the pH indicator is selected to change color around the midpoint (Methyl Red indicator in our implementation).

At the beginning of a computation, the $L_2$-PDA reactor contains deionized water and an aliquot of the $pH$ indicator. The processing of the symbols sequentially fed to the reactor leads to changes in pH whose value we assign to the stack. The $L_2$-PDA **Accepts** the input string if during the computation the *pH* ≥ midpoint-*pH but* is at midpoint-*pH* (empty stack) at the end of computation, that is, after adding "#." Conversely, the $L_2$-PDA *rejects* an input string if the *pH* falls below the midpoint-*pH* at any stage during computation (excess of ")", and attempting to "pop" from an already empty stack), or if the *pH* is larger than the midpoint-*pH* value at the end of computation (excess of "(", or the stack is "not empty") (cf. Fig. 2).

The response given by the $L_2$-PDA can again be interpreted in terms of a thermodynamic measure: the enthalpy yield of the computation $Y_{\Delta_H}$ (%). This is defined as the ratio between the enthalpy produced or consumed during computation divided by the total formation enthalpy of the chemical input:

$$Y_{\Delta H}(\%) = \frac{\text{reaction heat during computation}}{\text{formation heat of input string}} \times 100 \qquad (2)$$

$$Y_{\Delta H}(\%) = \frac{\sum_1^R \int_0^{t_{\text{end}}} v_i \, \Delta H^o_{r,i} dt}{\sum_1^n [j]_{\text{input}} \Delta H^o_{f,j}} \times 100 \qquad (3)$$

Here $R$ is the number of the reactions in the kinetic mechanism (at the level of coarse graining associated with the time between symbol processing), $v_i$ is the velocity of reaction $i$ (mol/($dm^3 \cdot s$)), $\Delta H^o_{r,i}$ represents the enthalpy of reaction $i$, $J$ is the number of symbols in the language alphabet, $[j]_{\text{input}}$ is the total change in concentration of species $j$ due to the input string, and $\Delta H^o_{f,j}$ is the formation enthalpy of chemical species $j$.

During computation, the dominant contribution to reaction heat occurs whenever a pair of parentheses is compensated via reaction $R_3$ (third reaction) in the mechanism (cf. Fig. 2): $OH^- + H^+ \rightarrow H_2O$. Hence the

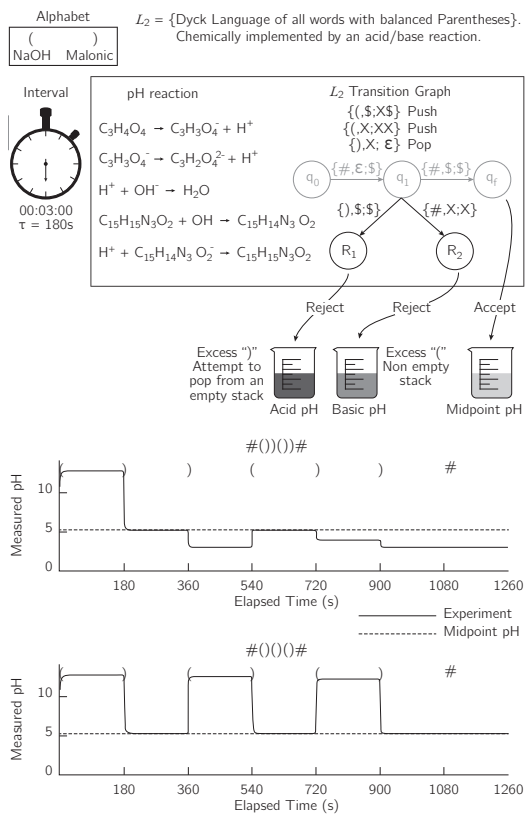

**Figure 2.** *Operation of a chemical one-stack pushdown automata: $L_2$ is recognized by a one-stack PDA. Here the reaction pH acts as the stack. If pH ≥ midpoint pH (intermediate gray tone or lightest gray tone, respectively) during computation, and pH = midpoint pH (lightest gray tone) at the end of computation, then the input string is accepted. Otherwise, if pH < midpoint pH (darkest gray tone) any time during computation, the string is rejected (attempting to "pop" from an empty stack). Also, if the pH > midpoint pH at the end of computation (intermediate gray tone), there is an excess of open parentheses and the string is rejected. Above are the experimental results for rejected ())()) and accepted ()()() words.*

enthalpy yield can be approximated as follows:

$$Y_{\Delta H}(\%) = \frac{\int_0^{tend} v \, \Delta H_r^o dt}{[\text{malonic}]_{\text{input}} \Delta H_{f,\text{malonic}} + [\text{OH}^-(aq)]_{\text{input}} \Delta H_{f,\text{OH}^-(aq)}} \times 100$$
$$\approx \frac{n_{\text{pairs}} c \, \Delta H_r^o}{[\text{malonic}]_{\text{input}} \Delta H_{f,\text{malonic}} + [\text{OH}^-(aq)]_{\text{input}} \Delta H_{f,\text{OH}^-(aq)}} \times 100$$

(4)

where $n_{\text{pairs}}$ is the number of pairs of parentheses that have been balanced and $c$ is the change in molarity of the solution due to the addition of each aliquot of malonic acid. For our *pH* reaction, the heat of the acid–base neutralization reaction is $\Delta H_r^o = -55.89$ kJ/mol, and the formation enthalpies can be found in standard thermodynamic databases (Haynes 2014).

By chemical engineering design of our 1-PDA, *Dyck* words maximize the enthalpy yield, whereas input strings that have excess of either open or closed parentheses will result in smaller enthalpy yields than strings with balanced parentheses. This again provides us with a thermodynamic metric to assess the result of the computation.

## *Chemical 2-Stack PDA/TM Recognizing Context-Sensitive Language* $L_3 = \{a^n b^n c^n$, *Where* $n > 0\}$

Finally, we demonstrate that our native chemical computing approach can be used successfully to recognize a language that only a Turing machine (TM) can recognize. A TM is an automaton equipped with an infinite tape and a read-write head working together with a finite set of transition rules. Initially, the input is written on the tape, each letter of the string written on one cell. The head can move left or right, reading and erasing and writing symbols on the tape based on the finite set of rules. As a part of the state transition, the TM decides if the next cell to be scanned is to the right or the left of the current scanned cell. The infiniteness of the tape and the possibility of moving either to the left or to the right of the tape are the factors that make the TM capable of recognizing all computable languages (Hopcroft, Motwani, and Ullman 2006; Turing 1936; Cohen 1991).

For our experimental implementation, we choose a well-defined and decidable language (Dueñas-Díez and Pérez-Mercader, Submitted 2018). The language $L_3 = \{a^n b^n c^n$, where $n > 0\}$, is made up of words consisting of n-repeats of **a**, followed by n-repeats of **b**, followed by n-repeats of **c**. Note that $L_3$ is a context-sensitive language not recognizable by either a FA or a one-stack PDA (Cohen 1991), and though it is not the most complex language a theoretical TM can recognize, it is quite convenient for an experimental implementation, as it brings into play all the features of a TM. Context-sensitive languages are recognized by a subclass of TMs, linearly bounded automata (LBA), in which only the cells occupied by the input are used for computation (Linz 2012).

TMs are equivalent to two-stack PDAs (Minsky 1961) because two stacks can emulate the function of moving right and left on an infinite (or arbitrarily long) tape. Taken together with the requirement of two interrelated stacks leads us to translate this into the chemical requirement of interrelated redox reactions and to oscillatory chemistry (Pérez-Mercader, Dueñas-Díez, and Case 2017; Dueñas-Díez and Pérez-Mercader, Submitted 2018). We have chosen arguably the best-known oscillatory chemistry, the Belousov–Zhabotinsky (Belousov 1959; Zhabotinsky 1964) reaction. As was the case before, alphabet symbols are carefully chosen to map into distinct pathways in the reaction mechanism and, consequently, to have distinct systematic effects on the measured oscillatory behavior (Dueñas-Díez and Pérez-Mercader, Submitted 2018): **a** is transcribed as an aliquot of sodium bromate affecting dominantly the autocatalytic production of $HBrO_2$ and catalyst oxidation, **b** is transcribed as an aliquot of malonic acid dominantly affecting the bromination of the weak acid and the reduction of catalyst, **c** is transcribed as an aliquot of NaOH affecting the pH-dominated subset of reactions, and **#** is transcribed as an aliquot of catalyst affecting the redox-dominated subset of reactions. The quan-titative recipes were selected and engineered to maintain the oscillatory regime for as long a word as possible, while simultaneously providing measurable changes in the oscillations. To implement this in a reactor, we used a combination of simulation and experimental studies.

The results of this experimental implementation have been reported in detail elsewhere (Dueñas-Díez and Pérez-Mercader, Submitted 2018). A key finding is that each state in the abstract TM transition graph has its own distinct chemical counterpart, for example, for our chosen language $L_3$, the reject due to the input containing *ba* has a different chemical signature than a reject due to an excess of *a*. There is a systematic clustering of chemical behaviors when mapping two basic phenomenological descriptors of the final oscillations (frequency and an oscillation amplitude-related difference measure). Experimentally, we find that words in the language are placed in a locus in this map, while rejected sequences (same as words, of course) lie either above or below (Dueñas-Díez and Pérez-Mercader, Submitted 2018).

To find a more intuitive criterion for acceptance/rejection, we introduce a metric based on the integral of the final oscillations (which we call the area) $A^{(\text{Word})}$ associated with the word undergoing processing in the computation:

$$A^{(\text{Word})} = V_{\max} \times \tau' - \int_{t_\#+30}^{t_\#+\tau} V_{\text{osc}}(t)\,dt, \tag{5}$$

where $t_\#$ is the time in reaction coordinates at which the end-of-expression symbol is added, $\tau'$ is the time interval between symbols minus 30 seconds (the first 30 seconds are discarded in the integration to allow for fast transients to dissipate), $V_{\max}$ is the maximum redox potential (all catalyst in oxidized form), and $V_{\text{osc}}$ is the measured redox potential, which can be well approximated by Nernst equation:

$$V_{\text{osc}} = V_0 + \frac{RT}{n_e F} \ln\left(\frac{[Ru(bpy)_3^{3+}]}{[Ru(bpy)_3^{2+}]}\right), \tag{6}$$

with $[Ru(bpy)_3^{2+}]$ and $[Ru(bpy)_3^{3+}]$, respectively, denoting the reduced and oxidized form of the catalyst, which can in turn be written in terms of the extent of reaction for the elementary redox reactions in the oxidation and reduction subsets as appropriately coarse grained. The quantity $n_e$ denotes the number of electrons involved in the reduction–oxidation process and is = 1 for this reaction. The redox potential is related to the Gibbs free energy $\Delta G$ (Kuhn and Försterling 2000) as

$$\Delta G_{\text{osc}} = -n_e F V_{\text{osc}}. \tag{7}$$

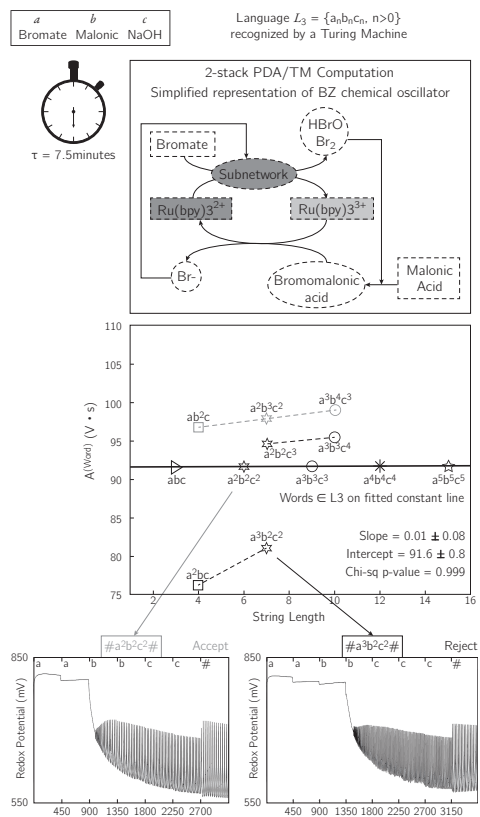

**Figure 3.** *A Belousov–Zhabotinsky-based chemical Turing machine for $L_3$ at work: $L_3$ is recognized by a two-stack PDA/TM, as shown by the constant area represented by the thick black line joining words in $L_3$ in the upper right-hand corner graph. Words not in $L_3$ lie elsewhere in the plot. During computation, if certain alphabet patterns in the redox potential V are detected, the strings are rejected. The rejection is specific for each type of reject. Here five words (in black font) were accepted and seven strings (in gray font) were rejected (two due to excess as, three due to excess bs, and two due to excess cs). The bottom panels compare the evolution of V for rejected $a^3b^2c^2$ (bottom right panel) and accepted $a^2b^2c^2$ (bottom left panel).*

We can thus rewrite the area defined above in terms of the Gibbs free energy corresponding to full oxidation $\Delta G'$ and the redox Gibbs free energy $\Delta G_{\text{osc}}$:

$$A^{(\text{Word})} = -\frac{1}{n_e F}\left(\Delta G' \times \tau' - \int_{t_\# + 30}^{t_\# + \tau} \Delta G_{\text{osc}}(t)\, dt\right). \tag{8}$$

The recipes for the alphabet aliquots can now be optimized to achieve a constant (i.e., n-independent or word length-independent) $A^{(\text{Word})}$ for words in $L_3$, while rejected sequences lie either above or below this value (cf. Fig. 3, top right). Hence, if the area $A^{(\text{Word})}$ is constant and independent of string length for the words in $L_3$, so is the integral of $\Delta G_{\text{osc}}$. Finally, we point out that the dimensions of this area are the same as those of the action in physics and that its origin reminds one of the mass-action law in chemistry.

## *Conclusions*

We have demonstrated experimentally that nonbiochemical chemistry in a homogeneous one-pot reactor, where the chemical inputs to be computed are fed sequentially at constant time intervals, has the capability to run successful computations at the three fundamental levels in the hierarchy of classical automata theory.

Our approach allows tailoring a chemical reactor to run a specific computation, that is, recognizing a specific language of interest, identifying an appropriate chemical transcription/translation of the alphabet and the chemistry of the automaton so that the reactor provides a distinctive thermodynamic/chemical response for those inputs that belong to said language. The design and operation of each chemical automaton follow similar principles, as the examples $L_1$, $L_2$, and $L_3$ illustrate. The elements of an automaton's tuple have chemical counterparts; for example, there are as many types of chemical "reject" states in the practical chemical automaton as "reject" states in the abstract automaton.

There are of course differences between abstract and actual automata. Any experimental chemical (or otherwise actual) realization of a TM necessarily has a noninfinite chemical tape; for that reason, it is most practical to implement the TM in the subclass of linear bounded automata. Note, however, that this is not too restrictive: by optimizing the operational strategy, including choice of reactor type and recipes, one can extend the tape to the needed length.

Computational versatility can also be enhanced by combining different chemical automata. The earlier discussed chemistries, including the Belousov–Zhabotinsky oscillatory chemistry, can be reconfigured to solve other languages of computational interest by appropriately selecting the alphabet symbols and their recipes such that all abstract-tuple elements have their chemical counterparts. The richness of time scales and nonlinearity in the Belousov–Zhabotinsky chemistry (or any other oscillatory chemistry) can be further exploited for computation. The reactions selected in this chapter are meant to provide illustrative examples for each class of automata. Other reactions in the appropriate classes can of course be used. Furthermore, other instances of specific computations can also be designed and carried out based on other specific chemistries, such as *pH* oscillators or biochemical oscillators. Finally, because abstract automata can be connected to create new automata (Hopcroft, Motwani, and Ullman 2006), we can imagine that chemical automata can likewise be interconnected to carry out more complex computations or generalizations of our automata (e.g., from a TM to a universal TM) if the underlying chemistries are compatible with each other, share common chemical species, and can be deployed in the same solvent media.

For each of the three implemented languages, we have identified a thermodynamic interpretation of the accept/reject states that is equivalent to the chemical response. Translating the criterion from purely chemical to its thermodynamic equivalent may simplify the interpretation of the acceptance/rejection of native chemical automata, as clearly seen for the case of the context-sensitive language $L_3$. In the examples discussed here, the thermodynamic interpretation involves thermodynamic potentials like enthalpy (languages $L_1$ and $L_2$) or Gibbs energy (language $L_3$). Such thermodynamic potentials were introduced in equilibrium thermodynamics to describe how closed systems approach equilibrium because, according to the extremum principles (Kondepudi and Prigogine 2014; Callen 1985), these potentials reach an optimal value at equilibrium. For example, in a closed system at constant pressure and temperature, the Gibbs free energy is at a minimum in equilibrium. Our chemical automata are open systems in nonequilibrium due to the semibatch feed of the chemical input, and hence these extremum principles do not apply. However, in the same way that an open system can be maintained at a (nonequilibrium) stationary state by the influx of matter and energy, we can direct the thermodynamic potentials to certain values and even to some (nonequilibrium) optimum values by feeding it with specific sequences in which matter and/or energy are inputs to the system. In this case, when the system reaches a (nonequilibrium) optimum, it is not as an (unavoidable) result from an extremum principle but driven or directed by our sequential chemical inputs of the language of interest and the specific recipes used. The chemical input directs the dominant reaction pathways, which in turn direct the thermodynamic pathways as well. This form of bootstrapping brings with it chemical control of the nonlinear, out-of-equilibrium chemistry itself.

Connecting the topology of complex reaction networks, their dynamics, and their thermodynamics is a recent and growing area of research (Rao and Esposito 2016). In our native chemical automata, the connection between chemistry and thermodynamics can contribute to better study and understanding of the energetic cost of computation, and prob-ably how to control this cost in the quest to approach Landauer's limit (Landauer 1961), at least for the important case of chemical automata. Furthermore, our approach does use liquid phase (dense) chemistry and kinetics and is not restricted by any approximations relying on the dilute gas approximation.

Nothing in the aforementioned precludes the extension of these results to biochemistry and biology. In particular, one can begin to think about the application of the preceding information-processing thermodynamics in the coming wave of new biochemistry-based oscillators (Novák and Tyson 2008), DNA-based oscillators (Srinivas et al. 2017), and computing ecologies of natural and synthetic bacteria—and finally, also to the study of the metabolic efficiency and the cost of chemical information processing and computation in extant living systems.